# High Resolution Observations of Cepheus A

V. A. Hughes[1], R. J. Cohen[2], and S. Garrington[2]

[1] *Astronomy Group, Department of Physics, Queen's University, Kingston, Ontario K7L 3N6*
[2] *University of Manchester, Nuffield Radio Astronomy Laboratories, Jodrell Bank, Macclesfield, Cheshire, U.K.*

**ABSTRACT**
New high resolution 6 cm observations have been made on Cepheus A using MERLIN, and combined with new VLA observations at 3.7 cm. Angular resolution with the latter was 200 mas, and with MERLIN was 60 mas, except for isolated unresolved sources where 33 mas was achieved. Unresolved objects at 60 mas were observed in Sources 2, 3, and in particular 9 which also was not resolved at 33 mas. There is no evidence for any other object as small as this with any significant flux density, although Source 8 was quiescent at the time. The upper limit to the size of Source 9 sets a minimum brightness temperature of $4.3 \times 10^5$ K, and adds credence to a previous suggestion that it is a gyrosynchrotron source. The compact objects of Sources 2 and 3 are thought to be produced by mass outflow from stars, which could be of spectral type B0 - B1, but this is uncertain. There is a discussion regarding the powerhouse for the molecular outflow. Sources 8 and 9, which are the highly time dependent objects, appear at the centre of the disruption of the high density gas, and their estimated high temperature of $10^7$ - $10^8$ K indicates that they could produce high velocity winds. On the other hand, the OH masers surrounding Source 2(ii) show an outward velocity of about 10 km s−1, which is small, but higher velocity winds could tunnel through adjacent spaces, and even be responsible for the 300 km s−1 bullet of Source 7. An extrapolation of the orthogonals to the IR polarization vectors are not accurate enough to pinpoint the source of the IR radiation, but it is estimated that Sources 2(ii) and 3(d)(ii) could supply the illumination. It is concluded that the two phenomena of molecular outflow and IR luminosity are not necessarily associated with one type of object only, but could be manifestations of two different types of object.

## 1 INTRODUCTION

Cepheus A has long been recognised as a star forming region, but details of its structure and the processes occurring there are only now being determined. It is situated in a CO condensation in a much larger CO cloud (Sargent 1977, 1979) at a distance of 730 pc, based on the distance to the nearby Cep OB3 association (Blaauw, Hiltner, & Johnson 1957). It is known to contain compact radio HII regions (Beichman et al. 1979; Rodriguez et al. 1980), OH and $H_2O$ masers (Blitz & Lada 1979; Wouterloot et al. 1980; Lada et al. 1981), and to be a strong infrared source (Koppenaal et al. 1979; Beichman, Becklin & Wynn-Williams 1979; Evans et al. 1981). In addition, the distribution of $NH_3$ (Brown et al. 1981) and of $HCO^+$ (Sandquist et al. 1982) have been observed, as well as the distribution and mass outflows in CO (Rodriguez, Ho & Moran 1983). These early observations had poor angular resolution, ranging from tens of arcsec for the molecular transitions, to at best 1 arcsec for some of the radio continuum measurements.

With the commissioning of the Very Large Array of the National Radio Astronomy Observatory[1], it was possible to resolve many more details of the region, and it was seen to consist of strings of apparent thermal radio sources (Hughes & Wouterloot 1984; Hughes 1985) such that if each were attributed to an HII region surrounding a star, then there were 13 - 15 B3 stars. At the time, this interpretation was somewhat suspect, since the apparent HII regions had dimensions of about 1,000 au, and under the then accepted scenario for star formation, unless there were some mechanism whereby the ionized gas could be contained, the HII regions would have dispersed in about 1,000 years. There was the added difficulty that it was unlikely that all 15 stars could be so close in their evolutionary status unless there were also some mechanism that triggered star formation. On the other hand, OH and $H_2O$ masers were associated with some of the HII regions (Cohen et al. 1984), and this has been taken as empirical evidence for the presence of stars. In addition, the total infrared emission was as expected from 15 B3 stars.

The possibility that the small size of the HII regions might cause them to be time dependent, led to a monitoring program using the VLA. Of interest was the discovery of Source 8, which was highly variable in a time interval of about 50 days (Hughes 1988), and also of a further variable Source 9 (Hughes 1991). In particular, the flux density of Source 9 was measured at four epochs, and its spectrum determined. The flux density peaked at about 6 cm, and an increase in flux density was accompanied by a movement of the peak flux to shorter wavelengths (Hughes 1991). The mechanism for such behaviour was attributed to gyrosynchrotron emission, and if this were so it would be produced in a region of size 1 au, temperature $10^7$ - $10^8$ K, and magnetic field of 100G. The density was about $10^6$ cm$^{-3}$.

The origin of the molecular outflows has been difficult to determine, principally because of the comparatively poor resolving power in the molecular line observations, in comparison with those in the radio continuum. This was demonstrated by Moriarty-Schieven, Snell & Hughes (1991) who showed the relative positions of the radio continuum, the 450 $\mu$m infrared, the CS emission, the CO outflows by Bally, Hayashi & Hayashi (1991), and NH$_3$ by Torrelles *et al.* (1985). However, the observation of an outward motion of the OH masers around Source 2, led to the suggestion by Migenes *et al.* (1992) that this source might contain the origin of the outflows. The later observations of a "bullet" moving at a transverse speed of 300 km s$^{-1}$, away from the region which contains also the Sources 9 and 2 (Hughes 1993), suggests that either source could initiate the outflows.

Since studies of the processes associated with Cepheus A may give important information on the physical processes which occur during the early stages of star formation, an attempt has been made to determine the properties of the region in more detail by making observations with higher angular resolution. This has been accomplished using the Multiple Element Radio Linked Interferometer (MERLIN)[2], which has a resolution of 60 mas at 6 cm but which can be reduced to 30 mas for point sources as is explained. The data have been combined with new data taken with the VLA at 3.7 cm, resolution of 200 mas, the latter being part of the continued monitoring of the region. This paper describes the above observations, in particular details of the four unresolved objects in the region with angular sizes < 60 mas (44 au). The observations are described in §2, while §3 describes the results. A general discussion follows in §4 and confirms the probable gyrosynchrotron nature of Source 9, describes the general nature of Sources 2, 3, 7, and 9, and the nature of the possible energy source for both the molecular outflow and the IR radiation. The conclusions are in §5.

## 2   THE OBSERVATIONS

### 2.1   The 3.7 cm VLA Observations

The 3.7 cm VLA observations were made 1992 November 25 in the A configuration, for a period of 3 hours. Angular resolution was ~ 200 mas and the bandwidth was 50 MHz. The data were edited and calibrated in the normal way, but due to a period of bad interference data on the calibrator 3C48 were badly corrupted. Since flux densities for the phase calibrator 2229+695 were considered to be unreliable, and the source could be variable, a calibration on 3C48 from an observing session on the same day, 20 hours previously, was used[3]. This was combined with the Cep A data, which gives a flux density for the secondary phase calibrator 2229+695 of 0.488 Jy. (It is of interest that previous measurements of the flux density of this source at 3.7 cm gave 0.95 Jy in 1986, and 0.754 Jy as observed by Patnaik *et al.* (1992).) We estimate that our derived values for flux densities in the Cep A region at 3.7 cm are accurate to within ± 3%.

The image was processed with MX, using uniform U-V weighting, and with one stage of ASCAL, using phase correction.

### 2.2   The 6 cm MERLIN Observations

Cepheus A was observed with 6 telescopes of MERLIN at 6 cm (4993 MHz) on 18 June 1992 for approximately 16 hours. The total bandwidth was 15 MHz (recorded in 1 MHz channels) in both left and right circular polarization. Amplitude calibration, band-pass calibration and correction for correlator offsets were done using a short observation of OQ208. The flux density of OQ208 was determined as 2.59 Jy relative to 3C286, whose flux density was taken from Baars *et al.* (1977) but reduced by 2% to allow for the slight resolution on the Mk2-Tabley baseline. A phase calibrator source 2300+638, selected from the MERLIN Calibrator List of Patnaik *et al.* (1992), was observed for approximately 2 minutes every 10 minutes. This source was mapped and self-calibrated and the amplitude and phase corrections applied to Cep A by interpolation. In order to maximize the sensitivity the data were weighted according to the telescope sensitivities: Fourier inversion and CLEANing produces a map with approximately 60 mas resolution and rms 80 $\mu$Jy/beam. The resolution can be increased at the cost of sensitivity by applying an additional inverse density weighting: this increases the resolution to 33 mas and the rms to 90 $\mu$Jy/beam.

The 'field-of-view' of these maps is limited by bandwidth smearing: at a radius of 9 arcsec a point source would be reduced in amplitude by 10%. To search for features beyond this radius a wide-field map was made by gridding the individual 1 MHz channels before Fourier transformation. In this map the field of view is limited by integration-time smearing to a radius of approximately 23 arcseconds.

### 2.3   Source Positions

The absolute positions of the features in the MERLIN map should be accurate to within about 15 mas: most of this error arises from the determination of the position of the phase calibrator source (Patnaik *et al.* 1992). This position was determined relative to a set of VLBI calibrator sources (Ma *et al.* 1990) using J2000 co-ordinates. The map coordinates have been precessed back to B1950 co-ordinates. The VLA positions were determined relative to the calibrator 2229+695. The position of this source has recently been measured as part of a VLBI calibration programme and the J2000 position given in the VLA list is accurate to within 1 mas. However, the B1950 position which was used for these observations appears to refer to epoch 1980. The fictitious proper motion between epoch 1950 and epoch 1980 is -0.0132s in RA and 0.122 arcsec in Dec. (in the sense 1980 minus 1950). In order to maintain consistency with previous VLA observations of Cepheus A, these offsets have been added to the MERLIN positions. Given their signal-to-noise ratio, the positions of individual components seen in the MERLIN map may be measured to about 10 mas relative to the map centre. In the VLA map the errors are similar, since the signal-to-noise is higher although the resolution is lower. Thus, the overall error in the positions of individual features is typically 20 mas in both the VLA and MERLIN maps.

### 2.4 Source Sizes

In the MERLIN 6 cm map of Cepheus A we detect several point-like components. In order to set upper limits to the sizes and hence to estimate brightness temperatures of the components we have investigated various techniques. For very weak sources, the most accurate technique is to shift the source to the phase centre and analyse the visibility data rather than the image. However, since Cepheus A contains several components this technique cannot be used here and the components must be analysed in the image plane. Fitting gaussian models to the components works well at high signal-to-noise and in order to assess the efficacy of this technique for very weak features we have performed simulations using sets of circular gaussian components with sizes ranging from 10 to 50 mas (FWHM). A flux density of 1 mJy was used for component sizes $\leq 30$ mas, but this was increased for the larger components to keep the peak brightness constant. The simulations were done using parameters appropriate to the MERLIN observations of Cep A and the maps produced had similar resolutions and rms values to the MERLIN maps. We found that using IMFIT to measure the component sizes the 10, 20 and 30 mas sources could not be distinguished. For 40 mas sources, there was a 50% success rate in resolving the sources, while for 50 mas sources the size could be measured to within 10 mas. For sources smaller than 30 mas uniformly weighted maps were no better than the naturally weighted maps in determining the component size.

In addition, we inspected the resultant beams obtained from AIPS, using a uniform weighting for the antennas, and a weighting according to their sensitivities, for both natural and uniform weighting across the array. The resulting beamwidths and the rms value over a region centred on Source 9 and comprising an area of $42 \times 42$ pixels ($60 \times 60$ mas) are also included in Table 1. An inspection of the $34 \times 33$ mas beam shows the presence of sidelobes as big as 30% of the main beam, but confined to within 68 mas from the centre of the beam. Inside 0.5 arcsec from the centre, sidelobes as big as 10% exist, but outside a radius of 2 arcsec sidelobes are negligible. Performing an FFT on the beam, it is seen that as expected the synthesized beam consists chiefly of the JB - Cambridge baseline of 220 km which produces a ring, with smaller diameter rings corresponding to baselines less than 70 km. The latter have the effect of reducing otherwise larger sidelobes. Thus, provided that a source is isolated by more than 2 arcsec from any neighbouring source, we believe that a resolution of about 30 mas can be achieved, but for source separations of less than 2 arcsec there may be uncertainties in flux density, although the CLEAN procedure will have the effect of reducing these by an amount dependent on the signal-to-noise ratio.

### 3 THE DATA

An inspection of the MERLIN data shows that the principal small diameter sources detected are Sources 2, 3(c), 3(d), and 9. At this time, Source 8 was quiescent. For reference purposes, we show in Figures 1 and 2 the central region of Cepheus A which contains these sources and which are, respectively, the 2 cm and 6 cm images of the central region of Cep A obtained 1990 May 8 using the VLA (Hughes 1991). These individual sources will be described separately, but we shall also include the VLA data on Source 7(c)(ii).

### 3.1 Source 2

Previous observations using the VLA, especially at 2 cm and resolution of 100 mas, suggested that Source 2 consisted of two individual objects (e.g. Hughes 1988), but each appeared to be time dependent. The new VLA observations at 3.7 cm, with resolution of 200 mas (145 au), but somewhat better sensitivity, are shown in Figure 3. There appear to be two chief components, possibly unresolved, together with a further two weaker ones. When observed with MERLIN at resolution of 60 mas, only one component is prominent, as shown in Figure 4.

IMFIT was used on the MERLIN and VLA data to obtain positions and flux densities for Sources 2(i) and 2(ii), and the data are tabulated in Table 2.

The only source common to the two data sets is 2(ii), whose position differs from the VLA position by 0˙.0074 in RA (52 mas) and 8 mas in Dec., which is not significant when note is taken of the somewhat complex MERLIN image. Using a resolution of 33 mas it appears that neither 2(i) nor 2(ii) has an appreciable component with a linear size as small as 24 au.

### 3.2 Source 3

The objects comprising Source 3 have been known to be complex, with the different components showing variability. Here we again compare the VLA 3.7 cm data with the 6 cm MERLIN data. Figure 5 shows the 3.7 cm VLA image, with the positions of the sources as determined using IMFIT shown as +'s. It is seen that Source 3(d) is resolved into two sources, which we label 3(d)(i), and 3(d)(ii), and an additional source is resolved between Sources 3(c) and 3(d). Figure 6 shows the corresponding 6 cm MERLIN image, with the above position of the 3.7 cm sources shown as +'s, corrected as for the case with Source 2. It appears that Source 3(d)(ii) is the only clearly unresolved object. Of interest is the presence of a number of components to Source 3(c), most of which are barely resolved, the most pronounced being the twin objects close to the centre of the VLA Source 3(c). The program IMSTAT was used on both the VLA and MERLIN data, and the results in terms of peak and integrated flux densities are shown in Table 2. Of note is the fact that the total flux density in the MERLIN data is somewhat greater than that in the VLA data, although this could be considered barely significant when errors are included. The chief point of interest is that Source 3(c) could actually consist of a number of discrete objects, each of size about 60 mas (44 au). The apparent decrease in flux density at shorter wavelengths could then either be the result of uncertainties in the data, or if real could be produced if the sources radiate by gyrosynchrotron radiation which peaks at 6 cm, as is the case with Source 9. However, unlike Source 9 there is no spectral data available, and no evidence for any component as small as 30 mas with flux density greater than 0.2 mJy.

**Table 1.** MERLIN and VLA Observations of Source 9.

| Array | Stat. Wt. | UV Wt. | Beam mas | Peak mJy | Integrated mJy | R.M.S. mJy | R.A. 22 54 + | Dec 61 45 + | S\N |
|---|---|---|---|---|---|---|---|---|---|
| MERLIN | Yes | Natural | 63×37 | 0.74 | 0.74 | 0.08 | 19.6935 | 45.666 | 9.3 |
|  | Yes | Uniform | 34×33 | 0.65 | 0.65 | 0.09 | 19.6935 | 45.666 | 7.3 |
|  | No | Natural | 52×49 | 0.81 | 0.81 | 0.10 | 19.6935 | 45.666 | 8.1 |
|  | No | Uniform | 37×37 | 0.82 | 0.82 | 0.13 | 19.6935 | 45.666 | 6.3 |
| VLA |  |  | 0.22×0.18 | 0.21 | 0.29 |  | 19.7059 | 45.653 |  |
| MERLIN-VLA |  |  |  |  |  |  | -0.011 | 0.020 |  |

**Figure 1.** VLA Image of the central region of Cepheus A at 2 cm, showing the different numbered sources. Contour levels are at -0.25, 0.25, 0.5, 1.0, 2.0, 3.0, and 4.0 mJy per beam. The image was obtained at 1990 May 8.

### 3.3 Source 7

The other source of particular interest is the "bullet", or Source 7(c)(ii) (Hughes 1993). It had previously been seen to be moving at a cross-speed of 300 km s$^{-1}$ in a direction away from the central region containing Source 2 and Source 9. It appeared to be compact at 6 cm with dimensions which over a ten year period had averaged 420 × 170 mas (300 × 120 au), an average peak flux density of 0.96 mJy, and integrated flux density of 1.50 mJy. The VLA image at 3.7 cm is shown in Figure 7, and data obtained using IMFIT is included in Table 2. It appears that there could be some small variation in flux density, but this is barely significant. Similarly, the dimensions of the source are not significantly different from those determined previously at 6 cm. The wide-field MERLIN map just detected 7(c)(ii), when a taper was applied to reduce the resolution to about 0.3 arcsec.

### 3.4 Source 9

Source 9 appears unresolved in the MERLIN map and from the simulations described above we place an upper limit on its size of 50 mas (FWHM). There is no significant difference in the position of this source as measured in the MERLIN and VLA maps.

However, since Source 9 is well separated from adjacent sources, we can apply the above analysis using the synthesized beams. The result of using IMFIT on the source for each of the beams is shown also in Table 1. The mean of the peak values is 0.755 mJy, with maximum variations about this +0.065 and -0.105, the latter being comparable to the rms noise in the area. We conclude that the maximum size of the source is 33 mas, or 24 au.

## 4 DISCUSSION

There are a number of aspects which arise from the determination of new upper limits of 60 mas to the size of Sources 2, and 3, and of 33 mas for Source 9. These can lead to a better understanding of the physical processes taking place, and possibly indicate the nature of the powerhouse(s) for the region.

### 4.1 Source 9.

**Figure 2.** VLA Image of the central region of Cepheus A at 6 cm, as for Figure 1. Contour levels are at -0.1, 0.1, 0.2, 0.4, 0.8, and at intervals of 0.4 to 4.4 mJy per beam. The image was obtained at 1990 May 8.

**Figure 3.** VLA image of Source 2 at 3.7 cm, and resolution 0.″2. Contour levels are at -0.05, 0.05, 0.1, 0.2, 0.4, 0.8, 1.6, and 3.2 mJy per beam. The image was obtained at 1992 Nov 25 in the A-configuration, angular resolution 0.″2. The +'s indicate the position of components 2(i) and 2(ii).

**Figure 4.** MERLIN image of Source 2, resolution of 0.″06. Contour levels are at -0.1, 0.1, 0.2, and intervals of 0.1 to 0.8 mJy per beam. The +'s indicate the positions of the components 2(i) and 2(ii) of Figure 3, corrected as described in the text.

The first object we consider is Source 9. The results of the analysis using different synthesized beams leads to the fact that it is not resolved at 33 mas, or 24 au. At epoch 1990 May 8, making a reasonable assumption from the shape of the spectrum that the source was optically thick at 20 cm, then $T_b D^2 \approx 2.5 \times 10^8$, where $T_b$ is the brightness temperature at this wavelength, and D(au) is the linear diameter (Hughes 1991). Assuming that there has been no appreciable change in size with time, if D = 24 au, $T_b \geq 4.3 \times 10^5$ K. However, since the spectrum shows that during times of activity as in the present case, the source becomes optically thin at wavelengths less than about 6 cm, the actual temperature at 3.7 cm, is probably a few percent greater than this. Thus, the estimated electron temperature is approaching the value of $T_b = 10^7 - 10^8$ K which is required

**Table 2.** Data on Sources 2, 3, and 7.

|  | Source | Beam arcsec | Peak mJy | Integrated mJy | R.A. 22 54 + | Dec 61 45 + | Source arcsec |
|---|---|---|---|---|---|---|---|
| VLA (3.7cm) | 2(i) | 0.22×0.18 | 1.33 | 1.94 | 19.0378 | 47.230 | 0.16×0.09 |
|  | 2(ii) | 0.22×0.18 | 3.66 | 6.57 | 19.0584 | 47.459 | 0.25×0.09 |
| MERLIN(6.0cm) | 2(ii) | 0.06×0.06 | 0.84 | 2.72 | 19.0639 | 47.341 | 0.10×0.08 |
| VLA (3.7cm) | 3(c) | 0.22×0.18 | 1.97 | 2.88 | 19.0274 | 43.989 | 0.17×0.06 |
|  |  |  | 0.71 | 1.46 | 19.1823 | 44.184 | 0.23×0.17 |
|  | 3(d)(i) |  | 1.17 | 1.94 | 19.2503 | 43.998 |  |
|  | 3(d)(ii) |  | 2.36 | 3.29 | 19.2683 | 43.961 | 0.13×0.12 |
| MERLIN(6.0cm) | 3(d)(ii) | 0.06×0.06 | 0.81 | 2.18 | 19.2676 | 43.849 | 0.09×0.06 |
| VLA (3.7cm) | 7(c)(ii) | 0.22×0.18 | 0.27 | 0.76 | 22.4234 | 35.160 | 0.37×0.16 |
| IMSTAT |  |  |  |  |  |  |  |
| VLA (3.7cm) | 3(c) |  | 1.89 | 3.16 |  |  |  |
| MERLIN(6.0cm) | 3(c) |  | 0.64 | 3.33 |  |  |  |

[R.A. and Dec are as determined for the respective images.]

**Figure 5.** VLA image of the Sources 3 at 3.7 cm. Contour levels are at -0.05, 0.05, 0.1, 0.2, 0.4, 0.8, and 1.6 mJy per beam. The +'s indicate those objects whose parameters are listed in Table 2.

for gyrosynchrotron radiation to be effective. Though this does not prove that the mechanism for the radiation is gyrosynchrotron, it is difficult to envisage any other acceptable mechanism.

As we have mentioned, during periods of flaring, there has been no indication of any increase in size, or change in position by more than 100 mas (73 au), and since the object has often been below the level of detection, it must have been in a period of activity during the MERLIN observations. However, there is no indication of emission around the object beyond about 33 mas.

A suggestion has been made that Source 9 could be the exciting source for the ionization in the line of Sources 7, and could be the driving source for the "bullet" (Hughes 1993), in which case it would likely be a Herbig Ae star, or a T-Tauri star. In this respect, the total radio luminosity at 6 cm, is about $10^{17.8}$ erg s$^{-1}$ Hz$^{-1}$, compared with an estimated $10^{16}$ - $10^{17}$ erg s$^{-1}$ Hz$^{-1}$ for post -T Tau stars (Güdel and Benz 1993). However, a higher radio luminosity might be expected since Source 9 is clearly flaring.

### 4.2 Sources 2 and 3.

**Figure 6.** MERLIN image of the Sources 3 at 6.0 cm. Contour levels are at -0.1, 0.1, 0.2, and at intervals of 0.1 to 0.7 mJy per beam. The +'s indicate the same objects as in Figure 5.

**Figure 7.** VLA image of the Sources 7 at 3.7 cm. Contour levels are at -0.05, 0.05, 0.1, 0.2, and 0.4 mJy per beam. The +'s indicate the position of the components in Hughes (1993).

Although Sources 8 and 9 are isolated, the unresolved components of Sources 2, and 3, are not. They appear to be embedded in much more diffuse emission. Previous determinations of spectral index $\alpha$ gave values of 0.6 and 0.4, respectively (Hughes 1985), where the flux density at wavelength $\lambda$ is given by $S_\lambda \propto \lambda^{-\alpha}$, and indicate that parts of the objects are optically thick, possibly the result of a stellar wind or mass loss. The present results as shown in Table 2 indicate that they are barely resolved at 60 mas (44 au), but there is no detectable flux density at a resolution of 33 mas, which is not unexpected for a beamwidth smaller in area by four times.

Source 2(ii) appears to have a size of 0.10 × 0.08 mas (73 × 58 au). Assuming an elliptical source, and using the expression $T_b ab = 3.71 \times 10^7 S_\nu$, where again $T_b$ is the brightness temperature, a(au) and b(au) are, respectively, the semi-major and semi-minor axes, and $S_\nu$ is the flux density in mJy, the brightness temperature is $8.9 \times 10^3$ K. This is somewhat greater than the normal electron temperature of a typical HII region, but we can use it to assume that the region is optically thick, and derive a lower limit for the electron density, N. If the optical depth, $\tau = 1$, then $N^2 L = 1.85 \times 10^{13}$ cm$^{-6}$ au, where L(au) is the physical depth of the region. Assuming a mean value for L of 66 au, then $N \gg 5.3 \times 10^5$ cm$^{-3}$. However, the fact that previous values of $\alpha = 0.6$ were obtained leads us to the conclusion that the electron density decreases inversely with the square of the radius, as would be expected for a star undergoing mass loss. A directivity and time dependence of the outflow could then account for some of the apparent variability seen around the radio image of Source 2. The presence of maser activity around the HII region would normally be taken as an indication that the contained star is young (e.g. Cohen 1989), and later we shall consider this further.

Assuming mass loss, then from the parameters for Source 2(ii) derived from the MERLIN observations, we can obtain some idea of the rate, and of the UV output from the contained star. We use equation (20) from the paper by Wright & Barlow (1975), from which we obtain a mass-loss

rate of

$$\dot{M}(H^+) = 2.18 \times 10^{-9} V_0 S_\nu^{3/4} \ M_\odot yr^{-1}, \quad (1)$$

where $V_0$ (km s$^{-1}$) is the speed of the wind at infinity, and $S_\nu$ (mJy) is the flux density. Assuming a value of $V_0$ = 300 km s$^{-1}$, equal to the transverse speed of the moving "bullet", then $\dot{M}(H^+) = 1.24 \times 10^{-6} \ M_\odot \ yr^{-1}$. The characteristic radius of the emitting region, $R(\nu)$, as derived from equation (11) of Wright & Barlow is then 30 au, reasonably consistent with the size of the source of 73 × 58 au determined using IMFIT. The number of UV photons required to maintain ionization of the wind, $N_{UV}$, is given from equation (23) of Wright & Barlow as $N_{UV} \gg 2.32 \times 10^{47} R_c^{-1}$, where $R_c(R_\odot)$ is the inner critical radius from where it is assumed the outflow originates. Assuming that $R_c$ = 5, $N_{UV} \gg 4.6 \times 10^{45}$ s$^{-1}$, equivalent to a B0.5 star of luminosity $10^4 \ L_\odot$ (Panagia 1972).

Source 3 contains a source very similar to 2(ii). It has the one comparatively large unresolved object, namely Source 3(d)(ii), although Source 3(c) could consist of a number of small unresolved objects. Source 3(d)(ii) does appear to have OH and H$_2$O masers associated with it (Cohen et al. 1984), and since its integrated flux density was measured at 2.18 mJy, its radio parameters are very similar to Source 2(ii), and we can apply the same arguments. However, there is no indication that it is responsible for outflows, except that the previously observed variability could be interpreted as local activity. Source 3(c) has a comparable flux density, but when the resolution is increased to 60 mas it appears to consist of a large number of smaller objects. We find it difficult to identify each of these as an HII region, rather with either shock excited ionized regions, or individual stars with associated ionization.

Thus, the presence of Sources 2(ii) and 3(d)(ii) could supply the total luminosity of $2.4 \times 10^{44} L_\odot$ observed in the IR (Evans et al. 1981). However, the above analysis assumes an initially neutral wind. If the outflow is the result of large temperature gradients in the chromosphere or corona, violent activity on the star, or other evolutionary processes, ionization will not be dependent entirely on photo-ionization, and in addition, a higher temperature will reduce the Hydrogen recombination rate. Under these conditions the UV flux required to produce the local ionization will be less, leading to a somewhat later type star than indicated. It has been noted, for example, that although mass loss from Mira variables appears to be driven by radiation pressure from a central star, the mass loss rates from some stars such as carbon stars, appear too high to be driven in this way (e.g. Knapp et al. 1982).

An alternative approach to determine the spectral type of the contained star can be made if we assume that most of the radio emission arises due to photo-ionization. For Source 2 and referring to Table 2, the total integrated radio flux density at 3.7 cm is $\sim$ 8.5 mJy, while that from the unresolved object is $\sim$ 2.72 mJy, indicating a high probability that most of the radiation is due to HII emission. If an HII region is optically thin, then it can be shown that the excitation parameter of the contained star, U(pc cm$^{-2}$), is given by

$$U = 1.44 \times 10^{-2} S^{1/3} D^{2/3}, \quad (2)$$

where S(mJy) is the flux density at 3.7 cm, and D(pc) is the distance to the HII region; the electron temperature is assumed to be $10^4$ K. At the wavelength of 6 cm, the constant term becomes $1.41 \times 10^{-2}$. The flux density at wavelengths less than 6 cm also appears to be increasing (Hughes 1985), but some of this could be due to radiation from the stellar wind, and to some radiation from surrounding dust. If we use the value for the flux density of Source 2 at 3.7 cm as 6.57 mJy, and a distance of D = 725 pc, then U = 4.1 pc cm$^{-2}$, which is equivalent to that for a ZAMS B1 star, for which the total luminosity is $5.2 \times 10^3 \ L_\odot$.

Thus, the estimated spectral types of the stars contained in Sources 2(ii) and 3(d)(ii), assuming ZAMS, lie between B0 and B1, though as discussed, there could be present ionization from stellar winds and associated shocks, which could indicate that they are stars of a somewhat later type.

### 4.3  Source 7.

The VLA observations on Source 7 are included not because any one of the objects is unresolved in the MERLIN 60 mas observations, but because Source 7(c)(ii) may have some association with the central objects of Cepheus A. Apart from Source 7(a), only the "bullet" is clearly detected at an integrated flux density of 1.67 mJy, and size 430 × 200 mas (300 × 145 au). Using similar arguments to those for Source 9, the brightness temperature is $2.1 \times 10^3$ K, consistent with the possibility that it is an optically thin thermal plasma. However, this may be too simplistic; rather it is a contained plasma of unknown temperature and thus density.

### 4.4  The Powerhouse.

From the above discussion, we infer that Source 9 is a star undergoing high activity. However, it is also noted that at the epoch of the MERLIN observations, Source 8 was below the level of detection, and since its behaviour in the past has been similar to that of Source 9, it is reasonable to assume that there exist at least two of this type of object in the region. Each of these might be expected to be the source of a high temperature wind. There are also at least two other stars in Sources 2(ii) and 3(d)(ii) which are undergoing mass loss and contributing to the overall ionization in the region. The additional presence of OH maser activity surrounding source 3(c), and an H$_2$O maser associated with Source 3(a) (Cohen et al. 1984), indicates the presence of further stars, but the latter has now expanded with a resultant decrease in flux density. Thus, there appear to be at least six stars in the region, but of somewhat uncertain spectral type. It is of interest that 4 ZAMS B1.5 stars could produce the luminosity and ionization in Cepheus A (Hughes & MacLeod 1993), but we suspect that the region is more complicated.

Additional information on the region comes from far-IR observations, which have been used to try and identify the source of the high luminosity. In particular, Lenzen, Hodapp & Solf (1984) obtained images at 2.2 $\mu$m, and though a number of IR objects were observed, only one, IRS 6(c), was in the immediate neighbourhood and this was a somewhat extended object. However, measurements of polarization showed the presence of scattering from what appears to be a reflection nebula. Drawing orthogonals to the polarization vectors gave an origin at a position close to Source 2, but with an error radius of about 2 arcsec. The interpretation was that Source 2 contained the high luminosity source. The relevant positions are shown in Figure 8, which shows not only the position of Sources 8 and 9, and of the masers in the region, but small squares show the suggested centroid of the source, P(1), and the position of IRS 6(c).

Also shown in Figure 8 is the centroid of the projected orthogonals to the polarization vectors by Joyce & Simon (1986), marked by the small square, P(2). The estimated uncertainty in position was 4 arcsec. If we consider both IR data sets, then it seems most likely that sources of the IR luminosity are Source 2 (most likely), and Sources 3(c), 3(d), and 8. It is also likely that any powerhouse for the molecular wind, whether or not it is the same as the source of the high IR luminosity, is time dependent, as suggested by Bally & Lane (1991) from their 2 $\mu$m observations.

It is also of interest to compare the positions of the contenders for the powerhouse with the recent $NH_3$ observations by Torrelles et al. (1993b). We show in Figure 9 the 20 cm image of the area on which we have sketched the three lowest contour levels of $NH_3$ emission. Source 9 is situated in a region of low emission which appears to be a cavity in a broken elongated region of emission, with an arm which extends to the north-east in the direction of the scattered IR radiation, and bordered by the Sources 4, 5 and 6, and a further arm which extends to the south-east at the end of which are situated the Sources 7, one of which is the "bullet" (Hughes 1993). The other contenders are seen against the $NH_3$, but in fact could be behind the denser gas.

The suggestion had been made that the $NH_3$ indicated either a dense slowly rotating disk (e.g. Torrelles et al. 1986), or an elongated feature (e.g. Moriarty-Schieven et al. 1991). From CO observations, Hayashi, Hasegawa & Kaifu (1988) had pin-pointed Source 2 as the source of most of the energetics of the region and the cause of the disruption of the high density disk of gas. However, the more detailed image by Torrelles et al. (1993a) shows Source 2 to be central to a smaller but even denser blob of $NH_3$ of size $3.3 \times 2.3$ arcsec, ($2400 \times 1700$ au), which is situated to the west of the cavity.

We can compare the size of the radio HII region and the area containing the OH masers with the size of the $NH_3$ condensation. We have plotted in Figure 10 the 3.7 cm contours, with the position of OH masers (Cohen et al. 1984) indicated by +'s. The overall dimensions of the radio HII regions are about $1.12 \times 0.45$ arcsec, or $800 \times 300$ au, while the diameter of the ring of OH masers is about 2 arcsec, or 1450 au (Cohen, Rowland & Blair 1984; Migenes et al. 1992). For comparison purposes, the $NH_3$ condensation has dimensions of $3.3 \times 2.3$ arcsec, or $2400 \times 1700$ au (Torrelles et al. 1993b). Although it is difficult to identify the radio source as sitting at the centre of the $NH_3$ blob, it seems that this is a reasonable interpretation and its extent is indicated by a large +. Thus it would appear that the HII regions are inside the blob of $NH_3$, and that the OH masers are towards the outside. The fact that there is an apparent secular decrease in the magnetic field associated with the masers (Cohen, Brebner & Potter 1990), and that they have an apparent outward motion, at a speed of $\sim 10$ km s$^{-1}$ (Migenes et al. 1992), lends credence to the fact that the region is expanding. To put a time-scale on the expansion, if the OH masers are now at a radius of 1 arcsec, and there is no squelching due to the expansion, then the age is 340 years. It has been noted by Habing & Israel (1979) that when the HII region expands to exceed 20 000 au, the conditions for OH masers disappear.

Regarding the high velocity CO outflow observed by (Bally & Lane 1991), Torrelles et al. (1993b) have compared it with their $NH_3$ contours. The resolution in each case is not sufficient to draw many conclusions, but it appears to be quadrupolar with directions close to those of the previously mentioned arms of the cavity.

## 5 CONCLUSIONS

New observations of Cepheus A at a resolution of 60 mas using MERLIN, have shown the presence of compact objects smaller in size than 44 au. In particular:

1. Source 9 is smaller than 24 au, which sets a lower limit to its temperature of $4.3 \times 10^5$ K, and tend to confirm a previous suggestion that it is a time dependent gyrosynchrotron source.

2. Source 2(ii) and 3(d)(ii) are barely resolved at a size of 44 au, but do not appear to have measurable components as small as 24 au. The interpretation is that they contain stars which are undergoing mass loss. Source 3(c) appears to break up into a number of objects of size 44 au, but these are unlikely to be individual HII regions. Rather they appear to be effects produced during the early stages of star formation.

It has not been possible to determine uniquely the source of the energetic molecular outflows, but possible contenders are objects contained in Sources 2(ii) and 3(d)(ii) which have appreciable mass loss, while Sources 8 and 9 are objects which are undergoing violent activity, have confirmed high temperatures estimated at $10^7$ - $10^8$ K, and thus highly likely to have strong stellar winds. It appears that the former could supply the IR luminosity for the region, which leads to the suggestion that the two phenomena of molecular outflow and IR luminosity may be manifestations of different objects.

## ACKNOWLEDGEMENTS

The authors wish to thank members of The National Radio Astronomy Observatory for advice pertaining to the operation of the VLA. Some of this work has been carried out under an Operating Grant from the Natural Sciences and Engineering Research Council of Canada.

[1] The National Radio Astronomy Observatory is operated by Associated Universities Inc., under cooperative agreement with the National Science Foundation.

[2] MERLIN is a UK national facility operated by the University of Manchester on behalf of SERC.

[3] The authors wish to thank A. Taylor and S. Daugherty for making available to them the antenna calibrations from 3C48.

## REFERENCES

**Figure 8.** VLA image of the central part of Cepheus A at 3.7 cm, showing the positions of Sources 8 and 9 and of the OH masers, indicated by +'s. The centroid of the polarization source determined by Lenzen, Hodapp & Solf (1984) is indicated by the small square P(1), and that of Joyce & Simon (1986) by P(2). Contour levels are at -0.05, 0.05, 0.1, 0.2, 0.4, 0.8, 1.6, and 3.2 mJy per beam.

**Figure 9.** VLA image of Cepheus A at 20 cm, showing also the first three contour intervals of $NH_3$ by Torrelles et al. (1993a).

**Figure 10.** VLA image of Source 2 at 3.7cm, with the position of OH masers shown by small +'s, and the extent of the $NH_3$ blob shown by the larger cross. Contour levels are at -0.05, 0.05 -0.1, 0.2, 0.4, 0.8, 1.6, and 3.2 mJy per beam.